\documentclass[%
 reprint,
superscriptaddress,
 amsmath,amssymb,
 aps,
]{revtex4-1}

\usepackage{color} 

\usepackage{graphicx}
\usepackage{caption}
\usepackage{subcaption}
\usepackage{dcolumn}
\usepackage{bm}

\begin{document}

\preprint{APS/123-QED}

\title{Adsorption of soft particles at fluid interfaces}

\author{Robert W. Style}
\email[]{style@maths.ox.ac.uk}
\affiliation{Mathematical Institute, University of Oxford, Oxford, OX1 3LB, UK}
\author{Lucio Isa}
\affiliation{Department of Materials, ETH Zurich, CH-8093 Zurich, Switzerland}

\author{Eric R. Dufresne}%
\affiliation{%
School of Engineering and Applied Sciences, Yale University, New Haven, CT 06520, USA 
}


%

\date{\today}

\begin{abstract}
Soft particles can be better emulsifiers than hard particles because they stretch at fluid interfaces.
This deformation can increase adsorption energies by orders of magnitude relative to rigid particles. 
The deformation of a particle at an interface is governed by a competition of bulk elasticity and surface tension.
When particles are partially wet by the two liquids, deformation is localized within a material-dependent distance $L$ from the contact line.
At the contact line, the particle morphology is given by a balance of surface tensions.
When the particle radius $R \ll L$, the particle adopts a lenticular shape identical to that of an adsorbed fluid droplet.  
Particle deformations can be elastic or plastic, depending on the relative values of the Young modulus, $E$, and yield stress, $\sigma_p$.
When surface tensions favour complete spreading of the particles at the interface, plastic deformation can lead to unusual fried-egg morphologies.
When deformable particles have surface properties that are very similar to one liquid phase, adsorption can be extremely sensitive to small changes of their affinity for the other liquid phase. 
These findings have implications for the adsorption of microgel particles at fluid interfaces and the performance of stimuli-responsive Pickering emulsions.  

\end{abstract}

\pacs{Valid PACS appear here}
\keywords{microgels, hydrogels, interfaces, adsorption, Pickering emulsions, colloids, elastic deformation, surface tensions, capillarity, elastocapillarity}
\maketitle

Emulsions are vital in many fields, including foods, cosmetics,  pharmaceuticals, and oil recovery \cite{brum06,levy90,cons95,schr92}.
Emulsions are typically stabilised by molecular surfactants, but they can also be stabilised with microparticles \cite{chev13,desh14}.
The resulting Pickering emulsions have many benefits over regular emulsions.
In particular, they are highly stable and avoid the use of potentially harmful or irritating surfactants \cite{chev13}.

A recent development is the use of soft colloidal particles (typically microgels) to make Pickering emulsions \cite{schm11,rich12,lyon12,desh14}.
These form easily \cite{cohi13,mont14,desh14}, and can be tuned \emph{in situ}, for instance by altering solvent properties, such as temperature and pH \cite{ngai05,ngai06,brug08,brug08b,tsuj08,mont14b}.
This has been demonstrated with emulsions created with poly(N-isopropylacrylamide) (pNIPAM) particles.
Since pNIPAM undergoes a reversible swelling/shrinking transition at a temperature close to body temperature, they have potential for 
 stimuli-responsive release of encapsulated active ingrediants for drug delivery \cite{desh14}.

A key difference between soft and hard particles as Pickering emulsifiers is that soft particles stretch as they adsorb  \cite{dest11,rich12}.
\begin{figure}
\centering
\includegraphics[width=7cm]{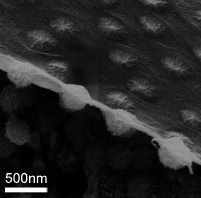}
  \caption{pNIPAM particles at an oil-water interface, imaged using cryo-SEM. The microgels were synthesized by free-radical precipitation polymerization with a 5 \% cross-linking density \cite{pina14} following standard routes \cite{pelt86,wu94,pelt00}.  
The cryo-SEM sample was prepared creating an interface between a 0.36 w/v \% aqueous suspension of the microgels and purified n-decane in a custom copper holder, letting the microgels spontaneously adsorb at the interface over 10 minutes. 
After shock freezing by liquid propane jets, the sample was fractured under high-vacuum and cryo conditions, removing the oil and exposing the particle-laden interface. 
A crack in the interfacial film (coated by a 3nm layer of tungsten for imaging) allowed a side view of the microgels adsorbed at the interface, where the aqueous phase was at the bottom of the field of view.  }
  \label{fig:lucio_ims}
\end{figure}
For example, microgel particles can have a much larger diameter at the interface than they do in the bulk. 
The extent of stretching depends on the cross-linking ratio of the hydrogel \cite{dest11,geis12,geis14}. 
Figure \ref{fig:lucio_ims} shows a side view of a p(NIPAM) microgel at a water/n-decane interface using cryo-SEM after freeze-fracture \cite{isa11}. 
While these soft particles have a spherical shape in solution, they are strongly deformed at the interface.
The central region of the particle remains somewhat spherical, especially the part exposed to the bulk water phase.
However, the particles are pulled strongly outward at the contact line, and a thin film of polymer spreads across the interface.

Here, we lay the groundwork for a theoretical understanding of the adsorption of soft particles at fluid-fluid interfaces.
In Section \ref{sec:limits}, we contrast the adsorption of rigid particles with the adsorption of fluid droplets.  
These limits bound the full range of behaviour for deformable particles.
We identify a critical point where deformable particles are extremely sensitive to small changes in the surface tensions.
Particles near this critical point could be ideally suited to making stimuli-responsive emulsions.
We then discuss two important scenarios for intermediate particle deformability.
In Section \ref{sec:elastic}, we calculate analytically the deformation  of linear-elastic particles that neutrally wet a fluid interface.
We find that elastic deformation is localized near the contact line, over a zone of width given by the ratio of the particle surface tension to Young's modulus, $\gamma_p/E$.
When the particle is much smaller than this lengthscale, it adopts a surface-energy-minimizing lenticular shape.
In Section \ref{sec:plastic}, we outline scaling arguments describing the plastic deformation of particles at a fluid interface.
Similar to the case of elastic deformations, we expect that plastic deformation should be localized over a zone near the contact line whose width is given by the ratio of the particle surface tension to its yield stress, $\sigma_p$.  
In the limit of large fluid-fluid surface tensions, particles can be driven far beyond the elastic limit and adopt unusual `fried-egg' morphologies.
We conclude by discussing the implications of our findings for microgel particles.

Throughout this paper, we make the following  simplifying assumptions.
First, we limit our attention to homogeneous spherical particles made of fluids or linear-elastic solids.
Second, we assume that surface tension of the particles is fluid-like.  
This means that the surface tension is isotropic and strain-independent, with numerical  equivalence of  the surface energy and surface stress.
This should be a good approximation for gels, including microgels, where the surface tension may be dominated by the properties of the embedded liquid phase \cite{hui13}.
Thus we will use the same symbol, $\gamma$, to refer to surface stress and surface energy throughout this paper.
Finally, we will simplify our notation by referring to the fluid-fluid interface as an oil-water interface, with the particles being initially dispersed in the aqueous phase.

\section{\label{sec:limits} Adsorption of fluid droplets and rigid particles}

\begin{figure*}
\centering
\includegraphics[width=15cm]{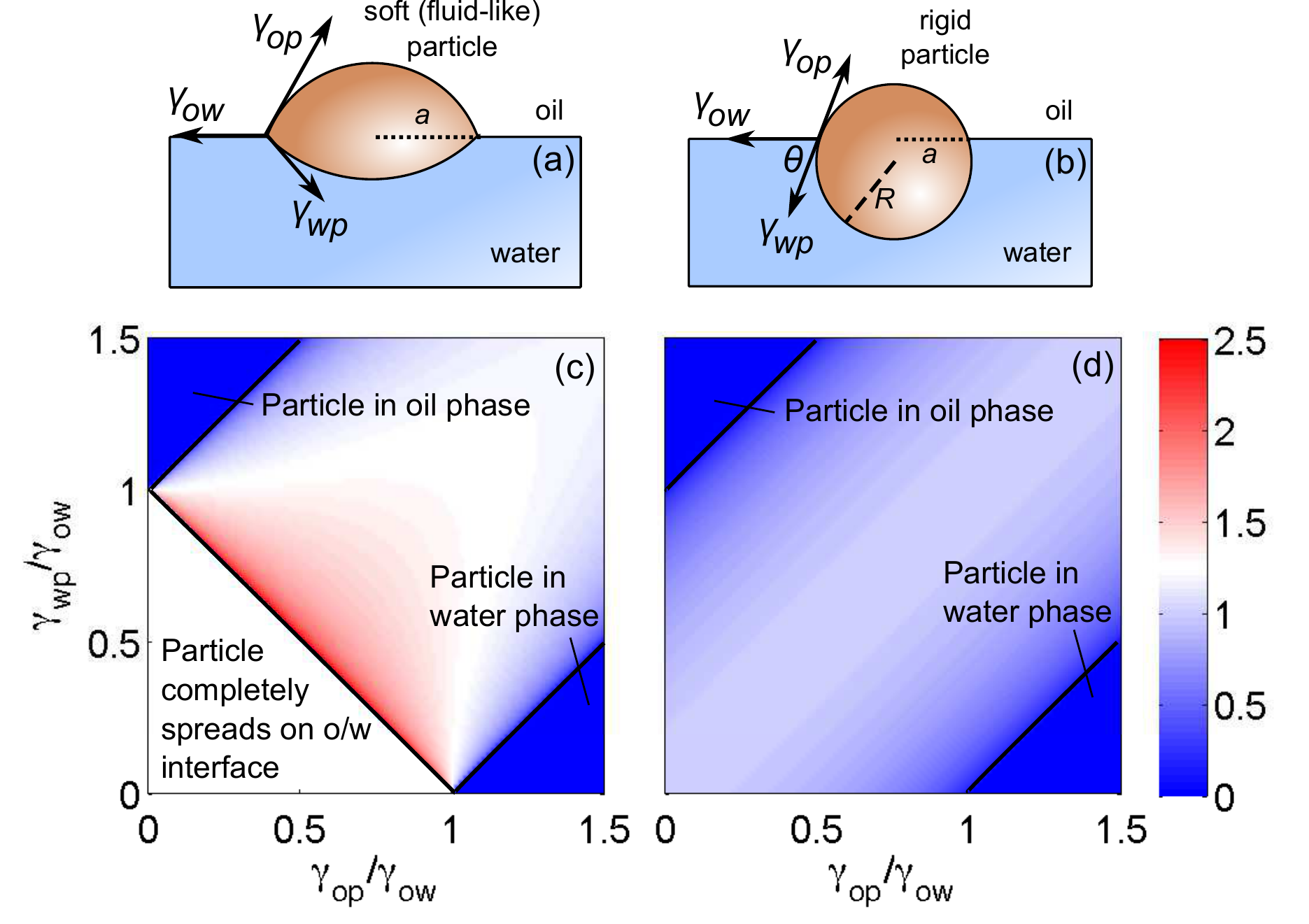}
  \caption{Limiting cases.  Schematic diagrams showing (a) perfectly-soft particles that take a lenticular form satisfying Neumann's triangle construction and (b) rigid particles obeying the Young-Dupr\'{e} relation.   Normalised contact radius $a/R$ for incompressible, initially-spherical (c) perfectly-soft, and (d) rigid particles.  Particles sit in the oil phase for $\gamma_{wp}/\gamma_{ow}>1+\gamma_{op}/\gamma_{ow}$, and in the water phase for $\gamma_{op}/\gamma_{ow}>1+\gamma_{wp}/\gamma_{ow}$. Perfectly-soft particles spread out at the oil-water interface for $\gamma_{op}+\gamma_{wp}<1$.}
  \label{fig:contact_radii}
\end{figure*}

The natural starting point is the two extreme cases of particle adsorption: (i) perfectly-soft particles (that behave like liquid droplets), and (ii) rigid particles.
These two cases will give us bounds on particle behaviour at interfaces.
In Section \ref{sec:elastic}, we show that particles will behave like one of these two extremes unless their undeformed radius, $R$, is near a characteristic material length scale.

In the limits of perfectly-soft and rigid particles, adsorption is completely determined by the surface tensions of the oil-water, oil-particle, and water-particle interfaces: $\gamma_{ow},\gamma_{op}$ and $\gamma_{wp}$ respectively.
Perfectly-soft particles take a lenticular shape, given by by Neumann's triangle construction \cite{dege10} which requires force balance at the contact line (Figure \ref{fig:contact_radii}a).
Rigid particles float so that they obey the Young-Dupr\'{e} relation: $\gamma_{ow}\cos\theta+\gamma_{wp}=\gamma_{op}$, where $\theta$ is the angle between the particle surface and oil-water interface on the water side (Figure \ref{fig:contact_radii}b).  

Soft particles can spread to cover up much more oil-water interface than hard particles.
We use Neumann's triangle and the Young-Dupr\'{e} relation to calculate the contact radii, $a$, of perfectly-soft, and rigid particles at interfaces respectively.
In both cases, $a/R$ only depends on $\gamma_{op}/\gamma_{ow}$ and $\gamma_{wp}/\gamma_{ow}$.
When $\gamma_{op}/\gamma_{ow}$ or $\gamma_{wp}/\gamma_{ow}$ are big, soft and rigid particles adsorb similarly at the interface, Figure \ref{fig:contact_radii}(c,d).
This is because large surface tensions $\gamma_{op},\gamma_{wp}$ keep soft particles approximately spherical -- like a hard particle.
However, when $\gamma_{op}/\gamma_{ow}$ and $\gamma_{wp}/\gamma_{ow}$ are small, $\gamma_{ow}$ can strongly deform soft particles, so they cover up far more interface than hard particles.
In particular, for $\gamma_{op}/\gamma_{ow}+\gamma_{wp}/\gamma_{ow}\leq 1$, perfectly-soft particles will completely spread, as indicated by the lower-left corner of Figure \ref{fig:contact_radii}(c).
This is analogous to complete wetting of a liquid on a rigid substrate.
In this limit,  surface tensions drive spreading that is ultimately limited by molecular-scale physics \cite{dege10}.
When $\gamma_{wp}/\gamma_{ow}>\gamma_{op}/\gamma_{ow}+1$ or when $\gamma_{wp}/\gamma_{ow}+1<\gamma_{op}/\gamma_{ow}$, particles do not adsorb to the interface, and instead sit completely in the oil/water phase respectively, regardless of their stiffness.
This is indicated by the upper-left and lower-right corners of Figures \ref{fig:contact_radii}(c,d)

Intriguingly, we find a critical point for perfectly-soft particles at $\gamma_{wp}/\gamma_{ow} = 0$ and $\gamma_{op}/\gamma_{ow} = 1$, see Figure \ref{fig:contact_radii}(c). 
Here, deformable particles are poised between complete spreading and total desorption.
Therefore, the adsorption of deformable particles lying near the critical point is extremely sensitive to small changes in surface tension. 
Such particles could be ideally suited for the formation of stimuli-responsive emulsions.

\begin{figure}
\centering
\includegraphics[width=9cm]{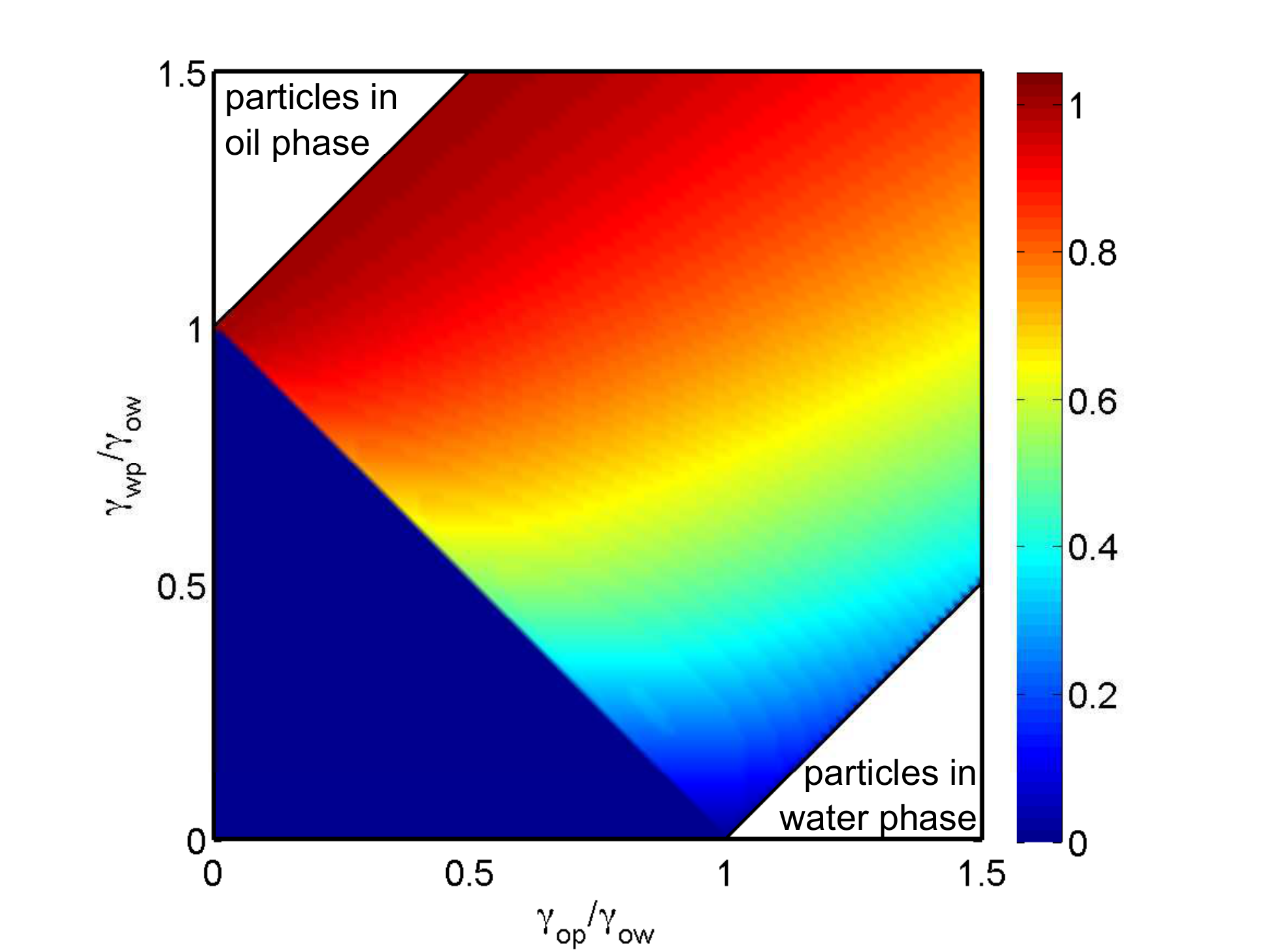}
  \caption{The ratio of rigid-particle over perfectly-soft-particle adsorption energies for particles adsorbing from water onto an oil-water interface. The adsorption energy depends sensitively on surface tensions when $\gamma_{wp}/\gamma_{ow} \ll 1$ and $\gamma_{op}/\gamma_{ow} \approx 1$.}
  \label{fig:ads_en}
\end{figure}

Deformable particles can bind much more tightly to the oil-water interface.
From the results above, we calculate the adsorption energy of particles from the water phase onto the oil-water interface: $E_{ad}\equiv -\Delta E=4\pi R^2 \gamma_{wp} + \pi a^2 \gamma_{ow} - A_w \gamma_{wp} - A_o \gamma_{op}$, where $A_w$ and $A_o$ are the final particle-oil/particle-water surface areas respectively.
Figure \ref{fig:ads_en} shows how the ratio of hard-particle to soft-particle adsorption energies $E_{ad}^h/E_{ad}^s$ depends on $\gamma_{op}/\gamma_{ow}$ and $\gamma_{wp}/\gamma_{ow}$.
For large $\gamma_{wp}/\gamma_{ow}$, adsorption of hard and soft particles is approximately the same:
in both cases, the particle mostly moves from the water phase into the oil phase, and $E_{ad}\sim 4\pi R^2 \gamma_{wp}$.
However, as $\gamma_{wp}/\gamma_{ow}$ gets smaller, soft particles are much more strongly adsorbed to the oil-water interface.
This is due to much greater spreading of the soft particles which covers up the oil-water interface.
When surface tension favours total spreading, $\gamma_{op}/\gamma_{ow}+\gamma_{wp}/\gamma_{ow}<1$, the adsorption energies of soft particles are arbitrarily larger than those of rigid particles (bottom-left corner of Figure \ref{fig:ads_en}).

\section{\label{sec:elastic} Elastic Deformation of Neutrally-Wetting Spheres }

In this Section, we quantify the deformation of linear-elastic particles to understand the transition from the perfectly-soft to rigid limits.
Previous work in related fields suggests that for particles with radii smaller than an elastocapillary length $L_{el}\sim O(\gamma_{op}/E,\gamma_{wp}/E)$, particle surface tensions will be the dominant force opposing stretching \cite{styl12,styl13c,styl15}.
Thus, such particles should adopt a lenticular shape like a perfectly-soft particle. 
On the other hand, for particles with $R\gg L_{el}$,  elasticity should be the dominant force opposing particle stretching \cite{styl12,styl13c,styl15}.
In this case, particle deformations should be small relative to the radius, so particles will appear `rigid'.
Thus we expect particles to appear either perfectly-soft or rigid unless $R\sim L_{el}$. 
Here, we show this rigorously by solving for the deformation of an elastic sphere at an interface, quantifying the transition from soft to rigid behaviour.
We choose the simplest problem of an initially-spherical elastic particle that neutrally wets an oil-water interface.
We assume that $\gamma_{op}=\gamma_{wp}=\gamma_p$, and that the particle is linear-elastic with Young modulus $E$, and Poisson ratio $\nu$.
We work in spherical co-ordinates $(r,\theta,\phi)$, with $\theta=\pi/2$ corresponding to the plane of the oil-water interface.

We calculate the particle's deformed shape by solving for the displacements in the particle $\mathbf{u}=(u_r,u_\theta,u_\phi)$, which satisfy the Navier equations:
\begin{equation}
(1-2\nu )\nabla^2 \mathbf{u}+\nabla(\nabla.\mathbf{u})=0.
\label{eqn:goveq}
\end{equation}
Stresses in the solid are given by
\begin{equation}
\sigma_{ij} + p_0 I_{ij}=\frac{E}{1+\nu}\left(\epsilon_{ij}+\frac{\nu}{1-2\nu}\epsilon_{kk}I_{ij}\right),
\label{eqn:stress_strain}
\end{equation}
where $I_{ij}$ is the Kronecker delta, and $\epsilon_{ij}=(\partial u_i/\partial x_j+\partial u_j/\partial x_i)/2$ is the strain.
$p_0=2\gamma_p/R$ is the Laplace pressure in the undeformed particle at zero strain caused by the particle's surface stress $\gamma_p$.
Note without $\gamma_p$ there is a strain singularity at the contact line \cite{jeri11}.

For boundary conditions we require that stress is bounded inside the particle, and that traction stresses at the particle's surface balance stresses caused by $\gamma_p$ and $\gamma_{ow}$:
\begin{equation}
\sigma.\mathbf{n}=\left[-\gamma_p {\cal K}+\frac{\gamma_{ow}}{r}\delta\left(\theta-\frac{\pi}{2}\right)\right]\mathbf{n}
\label{eqn:bc}
\end{equation}
(e.g. \cite{mora11,styl12,styl15b}). Here $\mathbf{n}$ and ${\cal K}$ are the normal and curvature of the surface respectively and $\delta$ is the Dirac delta function. Note that this boundary condition implies no shear stress at the particle surface.

We linearise the equations about the undeformed-particle state with $\mathbf{u}=0$ and $\sigma_{ij}=-(2\gamma_p/R)I_{ij}$ (e.g. \cite{styl15b}) to find that equation (\ref{eqn:bc}) becomes:
\begin{equation}
\sigma_{rr}=-p_0+\frac{\gamma_p}{R^2}\left(2 u_r +\cot\theta\frac{\partial u_r}{\partial \theta} +\frac{\partial^2 u_r}{\partial \theta^2} \right) +\frac{\gamma_{ow}}{R}\delta\left(\theta-\frac{\pi}{2}\right)
\label{eqn:linbca}
\end{equation}
and
\begin{equation}
\sigma_{r\theta}=0.
\label{eqn:linbc}
\end{equation}

We solve the governing equations (\ref{eqn:goveq},\ref{eqn:stress_strain},\ref{eqn:linbca},\ref{eqn:linbc}) using Legendre series (e.g. \cite{duan05}).
The general solution of equation (\ref{eqn:goveq}) inside the particle is:
\begin{equation}
u_r=a_0 r +\sum\limits_{n=1}^\infty \left(a_n r^{n+1}+b_n r^{n-1}\right)P_n(\cos\theta),
\label{eqn:ur}
\end{equation}
\begin{equation}
u_\theta=\sum\limits_{n=1}^\infty \left(\frac{a_n(5+n-4 \nu)}{(1+n)(-2+n+4\nu)}r^{n+1} +\frac{b_n}{n} r^{n-1}\right)\frac{d}{d\theta}P_n(\cos\theta)
\label{eqn:utheta}
\end{equation}
and $u_\phi=0$, where $P_n(x)$ are Legendre polynomials of order $n$.
Then applying equations (\ref{eqn:linbca},\ref{eqn:linbc}) and the identity
\begin{equation}
\delta\left(\theta - \frac{\pi}{2}\right)=\sum\limits_{n=0}^\infty \frac{2 n+1}{2}P_n(0) P_n(\cos\theta).
\label{eqn:deltafn}
\end{equation}
gives the solution for $n=0,1,...,\infty$:
\begin{widetext}
\begin{equation}
a_n=\frac{\gamma_{ow}}{4\gamma_p}\left(\frac{(n-2)(n+1)(2n+1)(1+\nu)(n-2+4\nu)P_n(0) R^{-n}}{2(n^2+n-2)(1+\nu)(1+2n^2(-1+\nu)-2\nu-n\nu) +\frac{RE}{\gamma_p}(2(1+\nu)+(n-1)n(-2-\nu+n(-3+2\nu))    ) }\right)
\label{eqn:an}
\end{equation}
\begin{equation}
b_n=-\frac{\gamma_{ow}}{4\gamma_p}\left(\frac{n^2(2n+1)(n+5-4\nu)(1+\nu)R^{-n+2}}{2(n^2+n-2)(1+\nu)(1+2n^2(-1+\nu)-2\nu-n\nu) +\frac{RE}{\gamma_p}(2(1+\nu)+(n-1)n(-2-\nu+n(-3+2\nu))    ) }\right).
\label{eqn:bn}
\end{equation}
\end{widetext}

\begin{figure*}
\centering
\includegraphics[width=13cm]{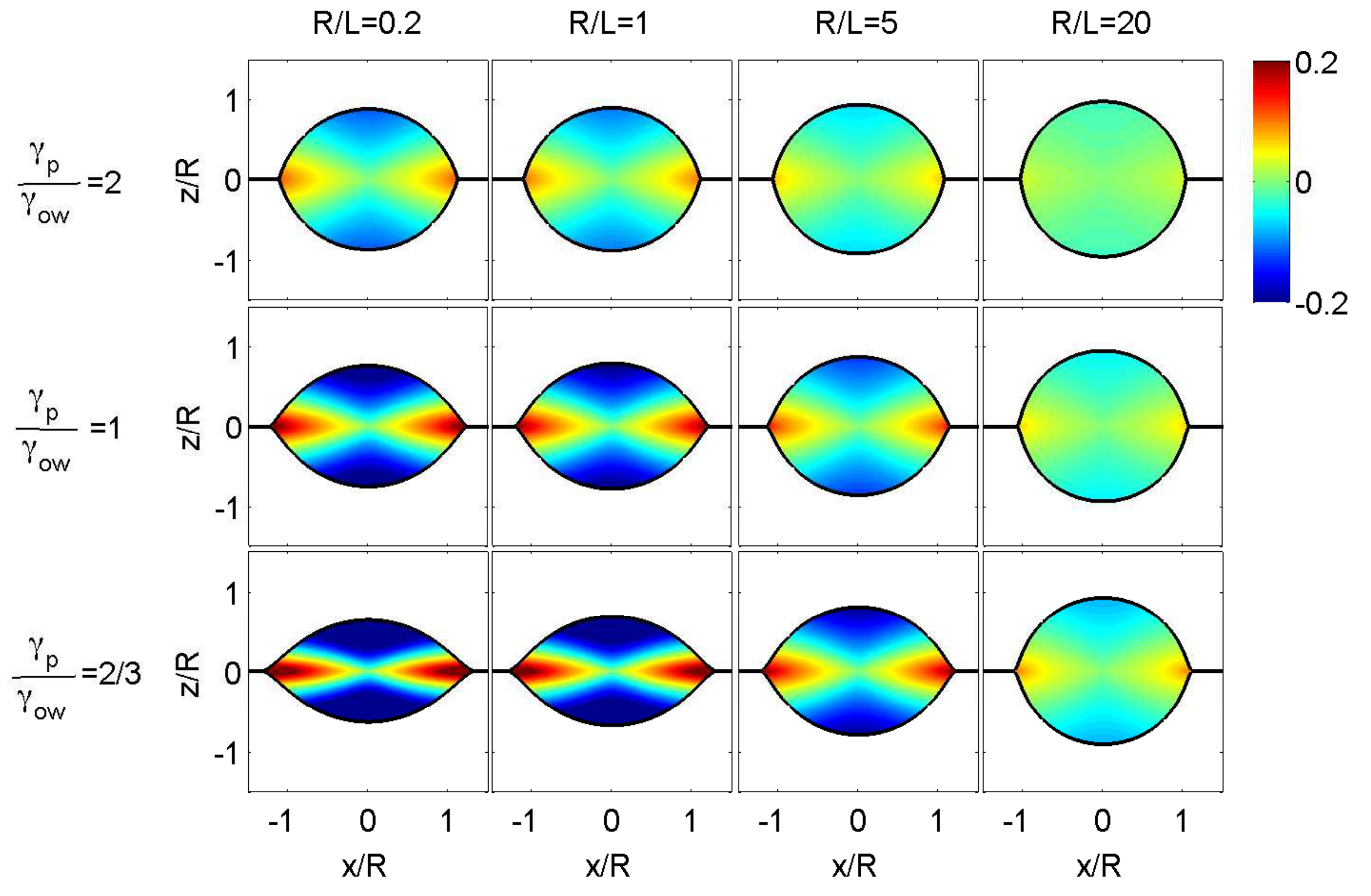}
  \caption{Deformation of soft particles at interfaces. Non-dimensionalised radial displacements $u_r/R$ for a range of $R/L$ and $\gamma_p/\gamma_{ow}$. Particles only show large-scale deformations when their radius is small compared to the elastocapillary length, $L$. For $R\ll L$, particles adopt a fixed, lenticular shape given by Neumann's construction.}
  \label{fig:part_disps}
\end{figure*}

These theoretical results allow us to investigate the parametric dependence of the deformation of adsorbed particles.
Equations (\ref{eqn:ur},\ref{eqn:utheta},\ref{eqn:an},\ref{eqn:bn}) show that $u_r/R$ and $u_\theta/R$ only depend on the two dimensionless parameters $R/L=RE/\gamma_p$ and $\gamma_p/\gamma_{ow}$.
This dependence is shown in Figure \ref{fig:part_disps} for $u_r/R$ with $\nu=1/2$.
When $R\gtrsim 20 L$, there is little particle deformation except in a small region at the contact line.
For smaller particles, deformations increase with decreasing particle size until, when $R\ll L$, the particle adopts a fixed lenticular shape that only depends on the ratio of surface tensions.
When $R\gg L$, changing $\gamma_p/\gamma_{ow}$ makes relatively little difference to the overall particle shape.
When the particle starts to stretch out ($R\lesssim 10 L$), decreasing $\gamma_p/\gamma_{ow}$ increases the particle aspect ratio.

The results in Figure \ref{fig:part_disps} illustrate the transition between `perfectly-soft' particle behaviour for $R\ll L$ and `rigid' particle behaviour for $R\gg L$.
This interpretation of the limiting scenarios can be understood by studying how equations (\ref{eqn:an},\ref{eqn:bn}) vary with $R/L$.
When $R\ll L$, the second term in the denominators drops out, and particle response is independent of $E$.
Then particle shape only depends on surface tension and particle compressibility, analogous to a droplet at the oil-water interface -- i.e. it behaves like an `perfectly-soft' particle.
On the other hand, when $R\gg L$, $a_n$ and $b_n$ are independent of $\gamma_p$ for small $n$.
Thus overall particle shape is unperturbed by adsorption, except near the contact line where small displacements should be $O(\gamma_{ow}/E)$ (e.g. \cite{mora10,styl13,styl13b,styl13c,mora13,wexl14}).

\begin{figure}
\centering
\includegraphics[width=9cm]{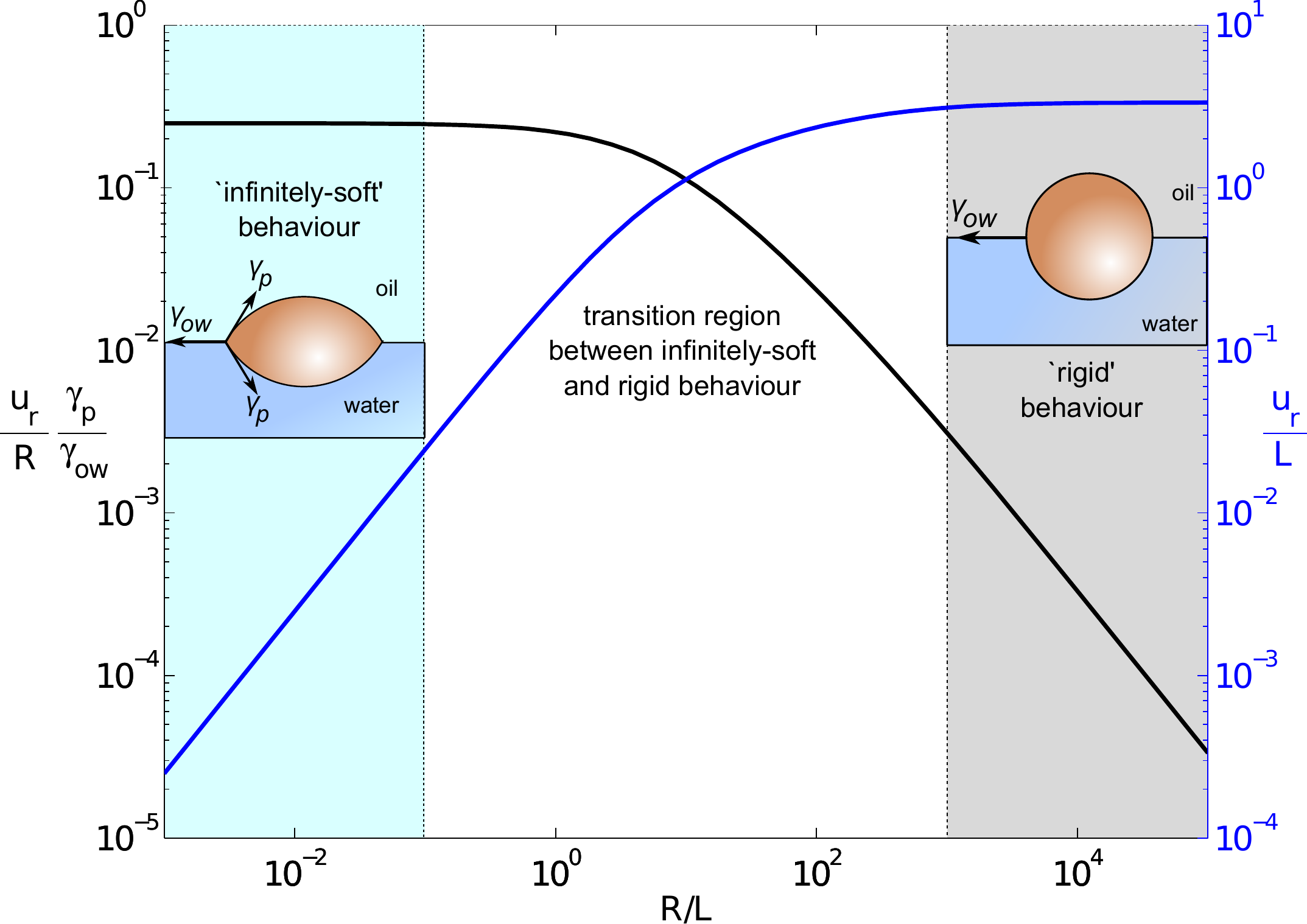}
  \caption{The scaling of contact-line displacements shows the transition from perfectly-soft to rigid behaviour of particles. The black curve shows ridge height $u_r(R,\pi/2)$ scaled by $R \gamma_{ow}/\gamma_p$. This plateaus for $R\lesssim 0.1L$ where particles take the lenticular shape predicted by Neumann's triangle construction (the perfectly-soft regime). The blue curve shows ridge height scaled by the elastocapillary length $L$. This plateaus for $R\gtrsim 10^3 L$. In this region $u_r/R\ll 1$, so displacements are negligible (the rigid regime). Particles only differ from the two limiting cases when $0.1 L\lesssim R\lesssim 10^3 L$.}
  \label{fig:urs}
\end{figure}

We can quantify when particles enter the `perfectly-soft' and `rigid' regimes by considering the elastic displacements at the contact line.
Figure \ref{fig:urs} shows how displacements at the particle waist $u_r(R,\pi/2)$ scaled by $R\gamma_{ow}/\gamma_p$ and $L$ depend on $R/L$ (for incompressible particles).
Small particles with $R\lesssim 0.1L$ have a rim of height $\propto R$. 
This is because the particle behaves like it is perfectly soft, and so takes the fixed, lenticular shape predicted by Neumann's triangle construction.
Large particles with $R\gtrsim 10^3 L$ have a rim of height $\sim O(L)$.
In this regime $u_r\ll R$, so particles are hardly deformed and appear rigid.
Only for $0.1 L\lesssim R\lesssim 10^3L$ do particles not behave as if they are perfectly-soft or rigid.

We can use the results above to calculate a particle's adsorption energy.
This comprises the changes in a particle's elastic energy and surface energy, and the change in surface energy of the fluid-fluid interface: 
\begin{equation}
\Delta E=\int_V \frac{1}{2} \epsilon_{ij} \sigma_{ij} dV + \gamma_p \Delta A -\gamma_{ow} \pi [R+u_r(R,\pi/2)]^2.
\label{eqn:deltaE}
\end{equation}
$\Delta A$ is the change in particle surface area, $A$, upon stretching. With the divergence theorem, the boundary condition (\ref{eqn:bc}), and force-equilibrium ($\nabla.\sigma=0$) the elastic energy term becomes:
\begin{equation}
\int_V \frac{1}{2} \epsilon_{ij} \sigma_{ij} dV=-\gamma_p\int_A \frac{{\cal K} \mathbf{u}.\mathbf{n}}{2}dA+\gamma_{ow} \pi R u_r(R,\pi/2).
\end{equation}
We insert this into equation (\ref{eqn:deltaE}), and note that 
\begin{equation}
\Delta A = \int_A\frac{{\cal K} \mathbf{u}.\mathbf{n}}{2}dA
\end{equation}
(e.g. \cite{cerm05}) to find the adsorption energy
\begin{equation}
E_{ad}=-\Delta E=\gamma_{ow}\pi R^2+\gamma_{ow} \pi R u_r(R,\pi/2).
\end{equation}
The first term is the adsorption energy of a rigid particle. 
The second term is the energy change due to particle stretching.

\begin{figure}
\centering
\includegraphics[width=9cm]{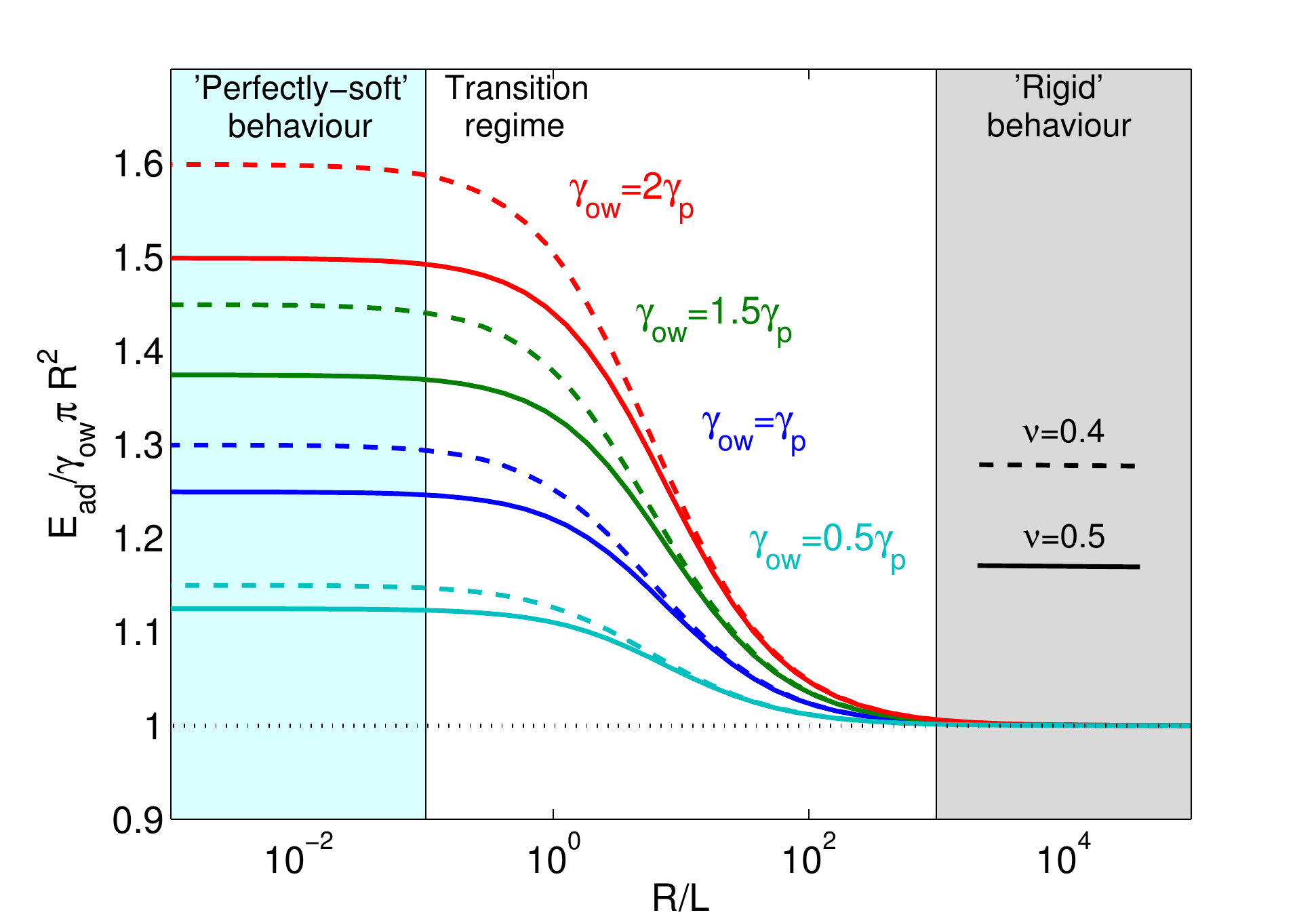}
  \caption{The adsorption energy of soft particles normalised by the adsorption energy of a rigid particle for various values of $\gamma_p/\gamma_{ow}$. Dashed/continuous curves correspond to $\nu=0.4/0.5$ respectively. Large particles with $R>10^3L$ appear rigid. Small particles with $R<0.1L$ have a constant adsorption energy, equivalent to that of a liquid droplet with the same surface tensions. Thus, they appear perfectly-soft. There is a smooth drop in the adsorption energy from the perfectly-soft to rigid limits.} 
  \label{fig:energy}
\end{figure}

Figure \ref{fig:energy} shows how $E_{ad}$ (normalised by the adsorption energy of a hard particle, $\gamma_{ow} \pi R^2$) depends on $R/L$ and $\gamma_p/\gamma_{ow}$ for the cases $\nu=0.4,0.5$.
As before, we see a transition from perfectly-soft ($R\lesssim 0.1L$) to rigid behaviour ($R\gtrsim 10^3L$) of particles.
Adsorption energy increases with decreasing $\gamma_p/\gamma_{ow}$, as particles stretch out and cover up more of the fluid interface.
Adsorption energy also increases with decreasing $\nu$ for the same reason.
In the perfectly-soft limit, $\gamma_{ow}$ cannot be greater than $2\gamma_p$ as surface tension favours the particle to spread completely at the oil-water interface, as in Figure \ref{fig:contact_radii}(a).
This limit poses  interesting questions for solid particles beyond the scope of this linear-elastic analysis and is discussed in the next Section.

\section{\label{sec:plastic} Plastic Deformation of Adsorbed Particles}

\begin{figure}
\centering
\includegraphics[width=5cm]{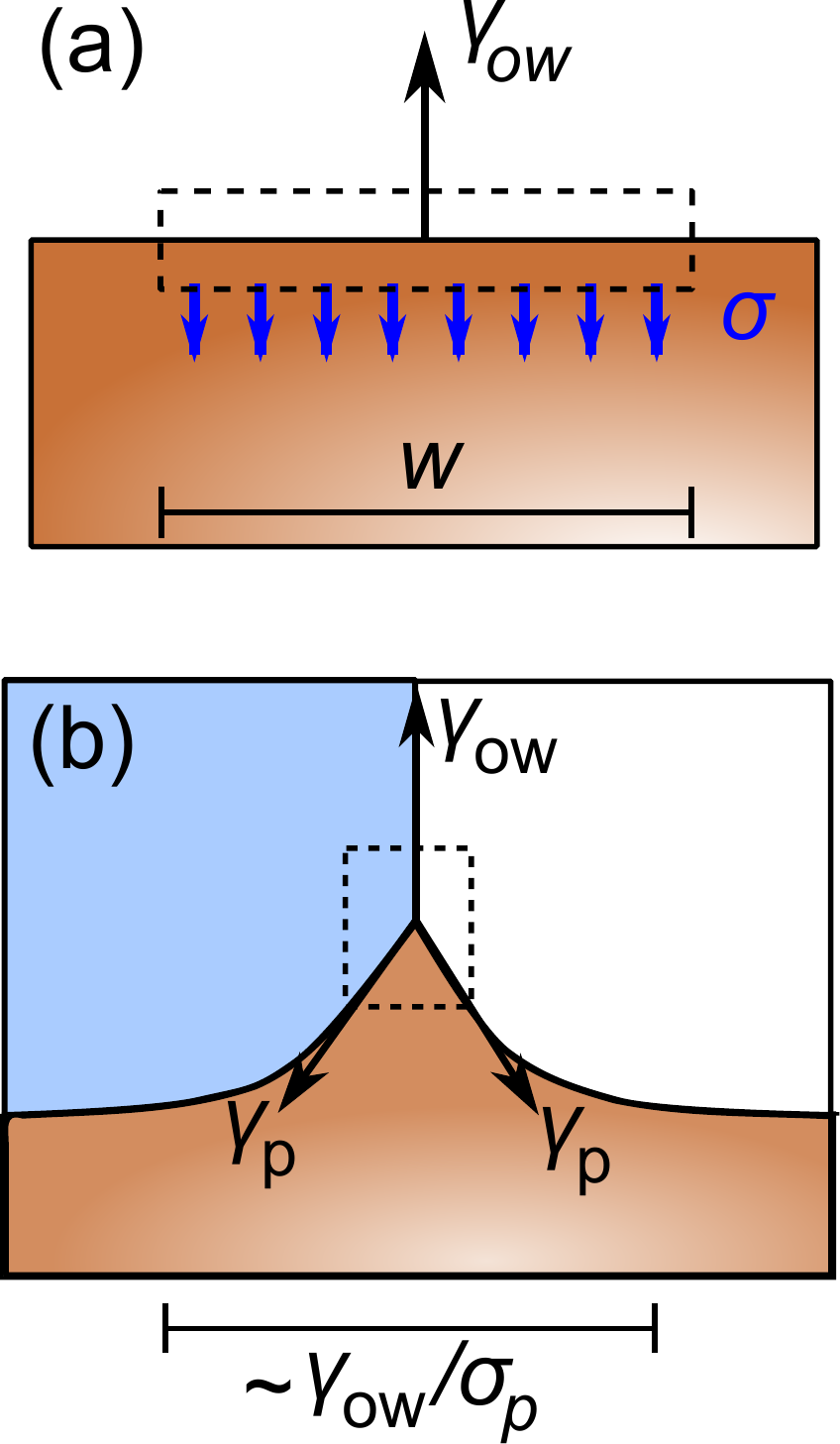}
  \caption{Deformation of a plastic surface by a contact line. (a) Control volume on an initially flat solid surface where surface tension, $\gamma_{ow}$, is balanced by an average bulk stress, $\sigma$, over a  region of size $w$. (b) Deep within the plastic region, indicated by the dashed box, the equilibrium ridge shape is determined by a Neumann balance between solid surface tension $\gamma_p$ and $\gamma_{ow}$ .}
  \label{fig:plastic}
\end{figure}

We now briefly discuss the plastic deformation of particles due to the surface tension of the fluid interface.
Consider a thin control volume of arbitrary width, $w$, near the contact line on an initially-flat surface, Figure \ref{fig:plastic}(a).
The control volume is acted on by the surface tension of the contact line, $\gamma_{ow}$, and the bulk stress, whose average is $\sigma$ over the width of the control volume.
Force balance requires that $\sigma w = \gamma_{ow}$.
The average stress within the control volume therefore diverges as $\sigma=\gamma_{ow}/w$.
Assuming that the particle has a yield stress $\sigma_p$,
there will be plastic deformation throughout the control volume when $w \ll L_{pl}=\gamma_{ow} / \sigma_p$ \cite{shan86}.
In other words, we expect a zone of plastic deformation of width $L_{pl}$ near the contact line.
Within this region, the surface tension will readily overcome bulk stresses and the equilibrium surface shape will determined by a  Neumann balance of surface tensions, as shown in Figure \ref{fig:plastic}(b) \cite{styl12,styl13}.
For an adsorbed spherical particle, the extent of plastic deformation  depends on the relative sizes of $R$ and $L_{pl}$.
When $R\ll L_{pl}$, the particle will appear `perfectly-soft', adopting a lenticular shape. 
When $R\gg L_{pl}$ the size of the plastic zone will be small in comparison to the particle size, localised to a plastic zone of size $L_{pl}$ near the contact line.

The transition between soft and rigid behaviour will be determined by the larger of the characteristic elastocapillary and plastocapillary length scales.
If $L_{pl} \gg L_{el}$, the deformation of adsorbed particles will be entirely plastic.
If $L_{pl} \ll L_{el}$, the deformation of adsorbed particles will be  reversible.
Adsorbed particles will appear to be perfectly-soft when $R \ll L$, where  $L=\max(L_{pl},L_{el})$.
When $R \gg L$, deformation will be localized near the contact line over a region of width $L$.
This deformation is negligible unless $\gamma_{ow}>2\gamma_p$.
In this limit, corresponding to the complete spreading discussed in Section \ref{sec:limits}, classical surface tensions cannot balance,
 leaving a divergent stress in the solid.
Unless $L_{pl}$ is comparable to or less than molecular dimensions \cite{hui14,lubb14}, this divergent stress will drive plastic flow of material out from the particle waist.
Surface tension favours the growth of this skirt to arbitary dimensions.
We expect its ultimate extent will be limited by the available surface area or molecular-scale physics.
We propose that this could contribute to the peculiar `fried-egg' morphologies observed in microgel particles at oil-water interfaces (Fig. \ref{fig:lucio_ims} and \cite{geis12,li13}).

\section{Conclusions}

We have considered the efficacy of soft particles as Pickering emulsifiers.
Since soft particles spread out at fluid-fluid interfaces, they can have much higher adsorption energies than hard particles.
Furthermore, in certain regions of phase-space, soft particles can be driven between complete spreading and desorption with extreme sensitivity to the surface tensions.
Whether a particle is `soft' or `hard' depends on the ratio of its size to elastocapillary and plastocapillary lengthscales.
Provided that the fluid-fluid surface tension is not too large, larger particles appear `rigid'.
Smaller particles appear `perfectly-soft' and behave similarly to liquid droplets at interfaces.
We have fully determined the transition between these two cases for neutrally wetting elastic spheres.
Our results suggest that plasticity becomes important when $\sigma_p \lesssim E$.
More generally, we expect \emph{plastocapillarity} to play an important role in a wide range of interfacial phenomena of solids where  elastocapillarity is known to be important, including  soft wetting \cite{shan86,styl12,styl13,styl13b,marc12,lima12,hui14}, surface flattening \cite{hui02,pare14}, instabilities \cite{mora10,mora14}, adhesion \cite{styl13c,cao14,sale13,xu14}, and composites \cite{styl15,styl15b}.

From a theoretical perspective, there are still many problems to be tackled.
Our analytic model is constrained to small deformations.  
However, we expect this to break down when the surface tensions favour complete spreading.
In this case,  nonlinear elasticity \cite{xu14,hena14,mora13}, or molecular dynamics modelling \cite{cao14} will be especially important.
It may also be interesting to explore what happens when the particle's surface energy is not equivalent to its surface stress.
Finally, an important question is how particles behave when they are no longer dilute, but packed (\emph{e.g.} hexagonally) at interfaces \cite{rich12,desh14b,pina14}.
Packed particles lose their spherical symmetry and do not fully stretch out on the interface -- this is expected to significantly affect their adsorption energy, and thus emulsion behaviour \cite{li13b}.

Experimental studies of soft particles at interfaces have focused on microgel particles.
What are the implications of our results on highly idealized particles for these real systems?
A number of limitations of our model are evident.  
First, our idealized particles are homogenous, while microgels typically have a gradient in crosslinking density, which may even include long dangling chains at the outmost layer.  
Second, real microgels are a two-phase system comprised of solvent and a swollen elastic network.
Our model does not consider the complex thermomechanical properties of such systems, but simply treats them as a one-phase system.
However, the two-phase nature of microgels does lead to one exciting implication of our work for real microgel particles:
Since the elastic network of a microgel particle is highly swollen by the solvent, the surface properties of microgels can be dominated by the solvent \cite{hui13}.
In that case, there is  little surface tension between the particle and the solvent (\emph{i.e.} $\gamma_{wp}/\gamma_{ow}$ approaches zero) and the surface tension of solvent against the other liquid phase would  be very close to the surface tension of the particle against the other liquid phase (\emph{i.e.} $\gamma_{op}/\gamma_{ow} \approx 1$).
This places microgel particles near the predicted critical point between complete spreading and desorption, as described in Section \ref{sec:limits} and shown in Figure  \ref{fig:contact_radii}(c). 
Thus, small changes in the affinity of the polymer for the oil phase due to changes in temperature or pH could potentially drive microgel particles between the extremes of complete spreading and desorption. 
This could be an ideal basis for a stimuli-responsive emulsion.
The precise location of a microgel particle on the phase diagram in Figure  \ref{fig:contact_radii}(c) will depend on the details of the polymer and solvent composition.
While common microgel systems may not already lie near the critical point, we argue that it may be advantageous to design future microgel systems to do so.

\section{Acknowledgements}

We thank Peter Howell and Joris Sprakel for helpful conversations. We thank Val\'{e}rie Ravaine and V\'{e}ronique Schmitt for providing the microgels and the ETH Zurich microscopy center ScopeM for access to instrumentation and support. We acknowledge funding from the National Science Foundation (CBET-1236086) for ERD, and from the Swiss National Science Foundation (PP00P2-144646/1) for LI.

\end{document}